# Identifying Biased Users in Online Social Networks to Enhance the Accuracy of Sentiment Analysis: A User Behavior-Based Approach

Amin Mahmoudi[1]

[1] Department of computing and decision sciences, Lingnan University, Hong Kong, aminmahmoudi@ln.edu.hk, corresponding author

*Abstract*—The development of an automatic way to extract user opinions about products, movies, and foods from online social network (OSN) interactions is among the main interests of sentiment analysis and opinion mining studies. Existing approaches in the sentiment analysis domain mostly do not discriminate the sentences of different types of users, even though some users are always negative and some are always positive. Thus, finding a way to identify these two types of user is significant because their attitudes can change the analysis of user reviews of businesses and products. Due to the complexity of natural language processing, pure text mining methods may lead to misunderstandings about the exact nature of the sentiments expressed in review text. In this study, we propose a neural network classifier to predict the presence of biased users on the basis of users' psychological behaviors. The identification of the psychological behaviors of users allows us to find overly positive and overly negative users and to categorize these users' attitudes regardless of the content of their review texts. The experiment result indicates that the biased users can be predicted based on user behavior at an accuracy rate of 89%, 67% and 81% for three different datasets

1. INTRODUCTION

1.1 Background

Sentiment analysis is a subfield of natural language processing, which is itself a subdomain of artificial intelligence. In sentiment analysis, the polarity of each sentence is extracted in order to categorize it as either positive, negative or neutral (Vishal A. Kharde & Sonawane, 2016). The idea of sentiment analysis arose from the limitation in human–computer interaction, where computers can process data and have a memory similar to humans, yet do not have the capability to understand human emotions. Although computers can process information more quickly and they have greater memory capacity than humans, they do not recognize sentiment and emotion. To overcome this problem, computer scientists have recently applied sentiment analysis and opinion mining with the aim of determining user behaviors in online social networks (OSNs) from users' psychological behaviors (Mahmoudi et al., 2019; Mahmoudi et al., 2021).

Nowadays, the recognition of the sentiments and opinions contained within users' review texts in OSNs plays a critical role in the success of marketing, recommender systems, and social events (Feldman, 2013; Da Silva et al., 2016; ). Recently, many researchers have introduced new techniques and methods to improve the accuracy of user sentiment identification in respect of specific topics in OSNs. They have analyzed the sentiments in comments at three levels: the document, the sentence, and the word (Hajmohammadi, Mohammad Sadegh et al., 2012). For each level, some techniques have been proposed, such as the support vector machine, naïve Bayes, and association rules (Singh & Shahid Husain, 2014). Corpus-, lexicon-, and dictionary-based methods can also be used for word sentiment classification. However, the majority of the methods in the literature only focus on the content without considering the characteristics of the users who create that content.

## 1.2 Motivation

It is very possible that we have all experienced situations in which we have come across people who are very positive or very negative in judging and commenting on products. On the other hand, because software systems calculate the average amount of sentiment about a product, there is a possibility that the opinions of these people will have a significant impact on that product. This problem has led to the emergence of affective computing, which aims to understand people's emotions through monitoring people's facial, vocal, and physical behaviors (Calabrese & Cannataro, 2016). Affective computing "relates to, arises from, or influences emotions" (Lisetti, 1998), whereas sentiment analysis tries to discover users' emotions through users' textual data. Although these two streams of research follow different approaches to identify users' emotions, they share a common goal, which is to determine users' emotions. It therefore follows that both of these two types of research may be able to identify users' psychological behaviors. In this study, we use a paramount hypothesis that is the basis of affective computing, namely, that human behaviors and emotions do not completely rely on laws and rules. This means that users' sentiments about businesses are not always based on logic, nor are they entirely emotional; rather, it can be said that users are sometimes biased and hold an extreme view that lies at either end of an opinion spectrum.

Some users are always negative and some are always positive. This means that some people are inherently very pessimistic and often make negative comments about businesses (restaurants, movies, etc.) or about other people because of their jobs, gender, race etc. In contrast, some people are inherently very optimistic and often make positive comments about all manner of things. In addition, user behaviors such as socialization, experience of, and lifespan in OSNs have a significant effect on user reviews in OSNs. Here, we consider those properties as user behavior, whereas the literal definition of behavior is "The way in which one acts or conducts oneself, especially towards others" (TKI, 2020). While being socialized is a social behavioral sign, the gaining of experience of or the spending of time in a social network (known as user lifespan in the OSN context) is a sign of social behavior.

## 1.3 Contribution

In this study, we would like to demonstrate how user characteristics as indicated by users' psychological behaviors have a direct effect on knowing user attitudes regardless of the content of users' review texts. The identification of biased users in OSNs is a very important issue for two main reasons. First, in sentiment analysis, the comments of biased users can have an adverse effect on the evaluation of a business. Hence it would be useful if this research were able to identify such users because their comments are not based on reality, but rather on their characteristics which are, in turn, rooted in their psyche. Second, biased users could also be considered as originators of racist and other prejudiced messages in OSNs. Thus, proposing a method that would be able to help to detect hatred and radical messages in OSNs would be of immense benefit. To the best of our knowledge, research has not been conducted elsewhere on this issue. Therefore, the outcomes of this study would be beneficial to society as a whole.

In this study, first, we analyze the role of user characteristics and behaviors in review texts and then show how to extract user attitudes based on users' psychological behaviors regardless of the content of the users' review texts. In this study and its experiments, it is hypothesized that it is more likely that people who are more socialized have a more positive attitude than those who are isolated. Similarly, people who are more experienced in terms of OSN usage will be more positive than those who are inexperienced. We can consider features such as socialization, experience etc. as user characteristics. To achieve its goal, this study uses the Yelp, COVID-19, and COVID-vaccine datasets. First, the important features that indicate user behavior are extracted from these datasets. Then, we compute the user sentiments in the three datasets. After that, the relationship between user sentiment and user behavior is illustrated. Second, the proposed model employs a new approach to make a neural network classifier to predict the presence of biased users based on users' psychological behavior in OSNs.

The rest of this paper is organized as follows: in the next section, the related work on this domain is reviewed. Then, the datasets used in this study are described in section 3. After that, the preliminaries and notations are provided in section 4. This is followed by section 5 in which the proposed model is described in detail. Then, in section 6, the results of some experiments are presented and the implications are discussed. Finally, in section 7, some conclusions are drawn.

2. RELATED WORKS

In the field of sentiment analysis, Jin and Zafarani (2018) demonstrated the possibility of predicting user sentiment through the analysis of the content in social networks. They used structural properties at four levels, namely, the ego, triad, community, and network level, to predict sentiment. In this way, they were able to analyze the effect of friends' sentiments on user sentiment. On the other hand, Tan et al. (2013) applied the concept of the social relationship to improve user-level sentiment analysis. The key assumption behind their idea was that users with mutual friendship are more likely to have similar opinions. Their study was conducted on Twitter data, and they focused on the user level rather than the tweet level because analysis at the tweet level is error-prone and time-consuming. In a similar vein, Yang and Eisenstein (2017) presented a method to solve language variation in sentiment analysis by using social attention. Also, Yang et al. (2016) used a sociological theory, namely, homophily, to overcome the ambiguity problem in Twitter text, which is an entity link system. In homophily theory, users who are connected are more likely to share similar behaviors and common interests. Meanwhile, Duyu Tang and Ting Liu (2015) used users and products influence on sentiment text because they claimed that existing studies mostly used text information. They introduced a model, which they named the user product neural network, to extract user and product information for use in the sentiment classification of documents. Gui et al. (2017) also leveraged user or product information as a feature with text information to train a sentiment classifier. To make heterogeneous networks, they considered four types of relation, namely, the word, word polarity, word user, and word product relations. In a later work, Gong and Wang (2018) used the self-consistency theory to discover user behavior in social media. They assumed that different users may share the same intents. Their proposed method consists of two main models: a logistic regression model to map the textual contents to the sentiment polarity, and a stochastic block model to capture the relatedness among users. Another novel method was introduced by Zou et al. (2018), where they based their model on social and topic content to identify user sentiment. Fornacciari et al. (2015) combined the social network structure with sentiment analysis to show how sentiment analysis could produce incorrect results depending on the network topology. In addition, they demonstrated that polarity of feeling can reveal the semantic connections in the network. Another related work worthy of mention in the context of this study is that by West et al. (2014), who used linguistic and social features to predict a person's opinion of another.

On the other hand, psychological studies have shown that people's characteristics have a significant effect on their attitude. Therefore, this study tries to use some social psychological characteristics such as socialization as a predictor of a positive attitude. This study tries to find this characteristic by using the OSN feature of number of friends because, according to ( Rubin and Bowker (2018) and Allan and Adams (2007), one of the main the criteria of being socialized is a person's number of friends. Conversely, this study supposes that asocial people (i.e., those with very few friends have a negative attitude in the OSN context. Moreover, the identification of relevant social factors such as societal norms, roles, and personal experiences would potentially be very useful and helpful for the success of our proposed model. Friendship is one of the most important factors for life satisfaction and has a significant effect on people's attitude (Berndt, 2002). The positivity of an individual can alter the attitude of their friends (Geoffrey, 2013), an outcome that occurs due to the presence of mirror neurons in the human brain. Moreover, as a user gains more experience, his or her positive attitude will be enhanced. Allport (1935) conceived attitude as "A mental and neural state of readiness, organized through experience" (p. 162). Fazio and Zanna (1981) conducted a series of studies to examine whether attitudes are formed through direct behavioral experience (vs. indirect non-behavioral experience) with the attitude object is a better predictor of people's subsequent behavior. Also, users who have long-term membership in OSNs gain more experience over the course of time, which also affects their attitude. Clearly, the demonstration of the existence of a relationship between the above-mentioned factors would not show causation. However, this study can show whether a type of user (i.e., highly socialized users) is overly positive. Any evidence of causation would need to be provided by a psychologist. Basically, our aim here is to try to find a relationship between sentiment and user behaviour in OSN datasets, and pave the way for psychologists to provide a theory about that relationship.

3. DATASET

This study uses the Yelp (*Yelp*, 2021), COVID-19, and COVID-vaccine datasets (Kaggle, 2020). Yelp is an OSN, in which users review the restaurants they have tried. The Yelp dataset contains six different JSON files: business, review, user, check-in, tip, and photo. The COVID-19 tweet dataset contains the tweets of users who applied the following hashtags: #coronavirus, #coronavirusoutbreak, #coronavirusPandemic, #covid19, and #covid_19. The dataset also includes tweets that incorporate the following additional hashtags: #epitwitter, #ihavecorona (Kaggle, 2020). The tweets were gathered for the period from March 4 to April 15, 2020. The COVID-19 tweet dataset contains approximately 18 million tweets. In addition, we used another dataset known as the COVID-vaccine dataset. The statistics for these three datasets is presented in Table 1.

Table 1. Dataset Statistics

| Dataset | No. of users | No. of businesses | No. of edges | No. of reviews | No. of check-ins | Time period |
|---|---|---|---|---|---|---|
| Yelp | 1,325,211 (104,924 users randomly selected) | 151,959 | 46,626,957 | 1,810,506 | 3,911,218 | 27/02/2004 to 11/12/2017 |
| COVID-19 | 553,848 | - | - | 17,900,881 | - | 04/03/2020 to 15/04/2020 |
| COVID-vaccine | 9,575 | - | - | 20,763 | - | 18/08/2020 to 17/11/2020 |

4. PRELIMINARIES AND NOTATIONS

*Definition 1: Online social network.* An OSN is an online service that facilitates communications among human users in a social network. Each OSN can be represented as $G\ (U, C, T)$, where U is a set of user IDs, $C$ is a set of triplets of the form of $(u_i, u_j, t)$, where $u_i, u_j \epsilon U$, $t \epsilon T$, and $T$ is a set of timestamps in which two users $u_i$ and $u_j$ communicate with each other and mostly T can be considered as the time of making a connection between two users. According to the above definition, an OSN is a dynamic domain in which the behavior of the network changes over time (Mahmoudi et al., 2020; Mahmoudi et al., 2020; Mahmoudi et al., 2018).

*Definition 2: Sentiment.* The Cambridge dictionary defines sentiment as "a thought, opinion, or idea based on a feeling about a situation, or a way of thinking about something" (Cambridge Dictionary, 2020). Sentiment analysis studies try to identify the emotions of humans from the review texts that they input via a computer or other Internet-enabled device. In this study, we measure the emotions of users in OSNs by computing the sentiment score of each review text (in Yelp) or tweet text (in COVID19 and COVID vaccine datasets) for any specific topic. For each user who tweets on a topic, a sentiment score vector is calculated and represented as $S_I = \{s_1, s_2, ..., s_n\}$, where $s_i$ represents the $i - th$ sentiment score of user $I$ who has tweeted on a specific topic. Each user has a number of tweets in the network. $TT_i = \{TT_1, TT_2, \ldots, TT_j\}$ represents a tweet on a topic by user *I*, as formulated in Equation (1):

$$S_I = Sentiment\ (TT_j)\ , if\ TT_j\ is\ tweeted\ by\ user\ I \tag{1}$$

Here, the sentiment is a function that computes the sentiment score of a tweet text, and the sentiment score can be a positive or negative value.

*Definition 3: Attitude.* Hendrickson Eagly and Chaiken (1993) defined attitude as a "psychological tendency that is expressed by evaluating a particular entity with some degree of favor or disfavor". Based on this definition, it can be said that people evaluate things through attitude (Haddock & Maio, 2008). This attitude can be positive or negative, and this can be said to be a normal way of evaluating something, but some people are overly positive or overly negative, which can be said to be an unnormal way of judging something. Based on this definition, this study assigns a value to each user attitude with respect to the degree of favor or disfavor that each user indicates in their review text (i.e., the sentiment score). The overall user attitude is formed by the sentiment of each user and can be represented as $A = \{a_1, a_2, \ldots, a_k\}$ for the OSN, where each $a_i$ is relevant to each user ($u_i$) in the OSN. Equation (2) is used to compute the attitude of each user. The summation of the user sentiment score over the number of reviews is considered as the user attitude as follows:

$$a_i = \sum_{idx=1}^{m} s_{idx} \tag{2}$$

where *m* is the number of reviews by user *I*; $a_i$ is the attitude of user *i*; and $s_{idx}$ is the *idx-th* sentiment value of the review text (comment) of user *i*.

We define user *i* as positive user if the user's attitude score is a positive value and vice versa, and it is represented by UA (user attitude) toward a specific business as shown in Equation (3):

$$UA_i = \begin{cases} positive & a_i > 0 \\ negative & a_i < 0 \end{cases} \tag{3}$$

*Definition 4 – Biased Users.* In this study, our definition of bias is related to the definition of attitude. This means that we consider a biased person to be someone who in most cases has an overly positive or overly negative attitude. Hahn and Harris (2014) described bias as "an inclination or prejudice for or against one person or group, especially in a way considered to be unfair". However, psychological studies do not define bias precisely and explicitly express a strong interest in something. Also, given that this is a completely new topic in the field of sentiment analysis, there is no precise definition of bias for this context. Therefore, in this study, we consider a statistical criterion based on the amount of sentiment value so that it can include a variety of definitions in the future. To do this, we obtain the sentiment distribution, that is, we see how much of the sentiment value is normal, and then we use standard deviation to determine the amount of sentiment that is abnormal. To illustrate, to calculate the sentiment distribution, first, we compute the mean value (μ) and standard deviation (σ), then we compute the distribution of sentiment with respect to (μ, σ).

Now, based on different amounts of standard deviation, different groups of biased users can be categorized. For instance, a sentiment value between μ + σ and μ − σ can be considered to indicate normal users, while a value bigger than μ + σ and smaller than µ − σ can be considered to denote biased users. In this study, we categorize biased users based on Equation (4) as follows:

$$U_i = \begin{cases} \text{overly positive} & a_i \geq \mu + 3\sigma \\ \text{overly negative} & a_i \leq \mu - 3\sigma \end{cases} \quad (4)$$

Meaning that $\mu - 3\sigma < U_i < \mu + 3\sigma$ denotes normal, otherwise biased, users.

Figures 1, 2, and 3 show the sentiment value distribution in the Yelp, COVID-vaccine, and COVID-19 datasets, respectively. As can be observed, the majority of users in all three datasets have a sentiment value in the range of the mean value and standard deviation, while a minority of users have a sentiment value bigger than the standard deviation and hence fall into the category of potentially biased users. Figure 1 shows the sentiment distribution in the Yelp dataset, where the mean value is 0.64 and the standard deviation is 1.75. In the Yelp dataset, around 97% of the users are normal, while 3% are overly positive because it can be seen that the sentiment value of the biased users is several times bigger than that of the ordinary users. Hence the sentiments of these users could affect the outcome of the judgment on a particular product, business or topic.

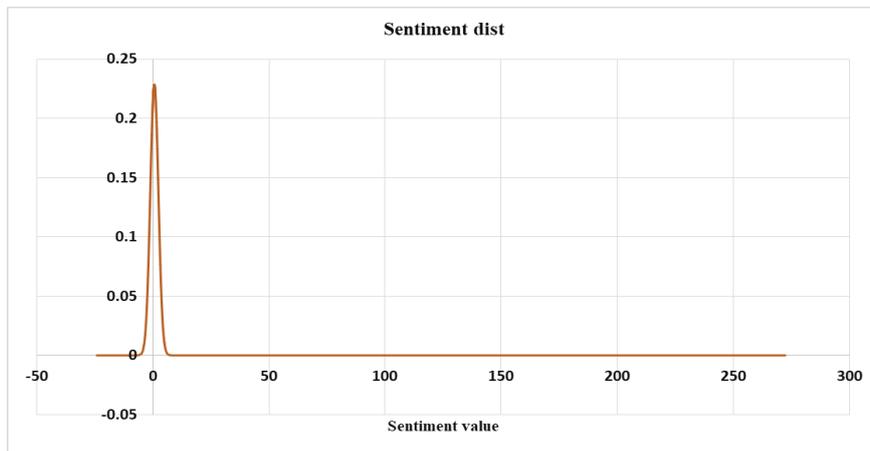

Figure 1. Sentiment value distribution of Yelp dataset

A similar pattern can also be seen in Figure 2, which provides the sentiment value distribution of the COVID-vaccine dataset. Figure 2 shows that the majority of users have a sentiment value of less than 1, but the sentiment value of some users is around 60, which means that the latter are overly positive. For the COVID-vaccine dataset, the mean sentiment value is 0.14 and the standard deviation is 0.9.

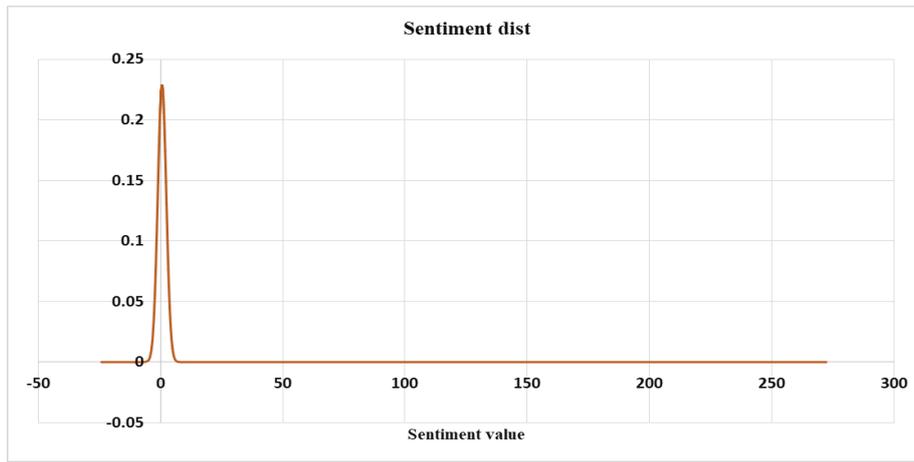

Figure 2. Sentiment value distribution of COVID-vaccine dataset

Likewise, Figure 3 shows that the pattern of the sentiment value distribution in the COVID-19 dataset is broadly similar. However, the amount of overly positive sentiment in this dataset is more prominent than the other two datasets, where the highest sentiment value is around 6500, while the mean value is 0.015 and the standard deviation is 10.49.

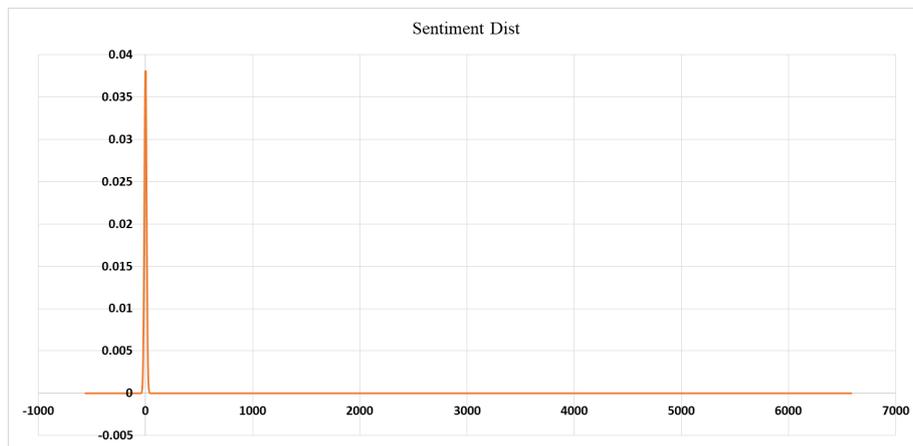

Figure 3. Sentiment value distribution of COVID-19 dataset

Table 2 describes the key notations used in the correlation matrix.

Table 2. Key Notations

| Notation | Description |
| --- | --- |
| NR | Number of reviews |
| LI | Length of time spent by user in the network (lifespan) |
| NFR | Number of friends |
| NFO | Number of followers |
| S_Score | Sentiment value |

5. PROPOSED METHOD

In order to meet the objectives of this study, first, we need to extract the significant features that affect users' sentiments about any topic. There are two ways to do this. The first is based on theory (theory-driven) and the second on data (data-driven). Then, by utilizing the influential features that have been extracted from user behavior, we make a neural network classifier. To compute the sentiment score, this study uses the sentiment package in R programming language. Specifically, we use the "sentiment_by" function of this package. The output of this function is a data frame that contains the "avg_sentiment" which shows the sentiment (polarity) score.

5.1 Theory and Hypothesis Development

To analyze the role of user characteristics in review texts, first, we extracted the user characteristics available in the datasets listed in Table 1. These characteristics included the users' number of friends, number of followers, lifespan in the OSNs, and the number of reviews they had made. Then, the relationship among these attributes and the presence of biased UAs were identified and utilized as the basis for a new model of sentiment analysis based on user behavior. To compute the scores of user review texts, the RSentiment package in R programming language was used.

The development of a method to identify biased users based on user characteristics (psychological behavior of users) was the main objective of this study, To achieve this goal, first, we computed the sentiment score of each user review. Then, based on the standard deviation of their sentiment scores, a criterion was defined to specify biased users. After that, we created a neural network classifier model based on the significant user characteristics that are the signs of users' psychological behavior. Finally, we tested the developed model to determine whether it was able to detect biased users correctly or not.

One of the key parts of the method was feature selection (of user characteristics – psychological behavior). In this study, we used user texts to extract these features. One of the main tools that can be used to select significant features is the correlation matrix. This matrix shows how features are associated with each other. In this study, we tried to find the amount of association between users' characteristics and users' sentiment values. The strength of the correlation is indicated by the correlation coefficient. There are two main types of correlation coefficient, namely, Spearman and Pearson (Schober et al., 2018). The tweet datasets and Yelp dataset that we employed in this study contained verified and sentiment (emotion) variables, which are types of ordinal data. The Spearman's rank correlation is suitable for such data, therefore we used it to measure the association between the above-mentioned user characteristics and their sentiment scores.

*Correlation coefficient.* The correlation coefficient r can be computed as follows: Suppose that there are two variables $R$ and $S$, and that each one consists of some values $R_1, R_2, R_3 ..., R_n$ and $S_1, S_2, S_3 ..., S_n$, respectively. Let the mean of $R$ be $\bar{R}$ and the mean of $S$ be $\bar{S}$. Hence $r$ is:

$$r = \frac{\sum(R_i-\bar{R})(S_i-\bar{S})}{\sqrt{(R_i-\bar{R})^2(S_i-\bar{S})^2}} \tag{5}$$

Figure 4 shows that value of r for different amounts of association between sample values. In Figure 4, (a) show a strong negative correlation, (b) shows a weak negative correlation, (c) shows a very weak (negligible) negative correlation, (d) shows a strong positive correlation, (e) shows a weak positive correlation, and (f) shows a very weak (negligible) positive correlation.

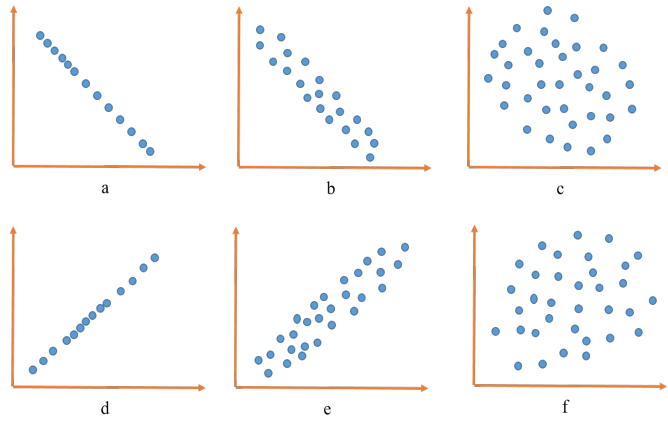

Figure 4. Correlation coefficient *r* between two variables

The correlation coefficient *r* is a summary measure that describes the extent of the statistical relationship between two variables. The value of a correlation coefficient can range between −1 and +1, and there is a greater association when the value is +1 or −1, indicating an exact association. A correlation coefficient value close to +1 indicates a positive relationship between two variables and a value close to -1 shows a negative relationship between two variables. A correlation coefficient of 0 would indicate that there is no association between the two variables.

According to this study's definition of biased users, we computed the mean value of the sentiment scores to classify users into negative and positive users. In addition, we defined different ranges of sentiment score with regard to the mean value to show the strength of bias. Thus, the mean value is the indicator in this study. In other words, our analysis of the correlation and relationship between user sentiment and the selected user characteristics is based on this indicator. Table 3 shows the mean value of the three datasets used in this study.

Table 3. Sentiment Value of Datasets

| Dataset | Sentiment mean value | Mean value of positive sentiment | Mean value of negative sentiment |
|---|---|---|---|
| Yelp | 0.64 | 0.76 | -0.18 |
| COVID-19 | 0.015 | 0.625 | -0.72 |
| COVID-vaccine | 0.137 | 0.322 | -0.204 |

Figures 5, 6, and 7 show the correlation matrix of the Yelp, COVID-19, and COVID-vaccine datasets, respectively. The relationship is provided based on the available user characteristics (features) in each dataset.

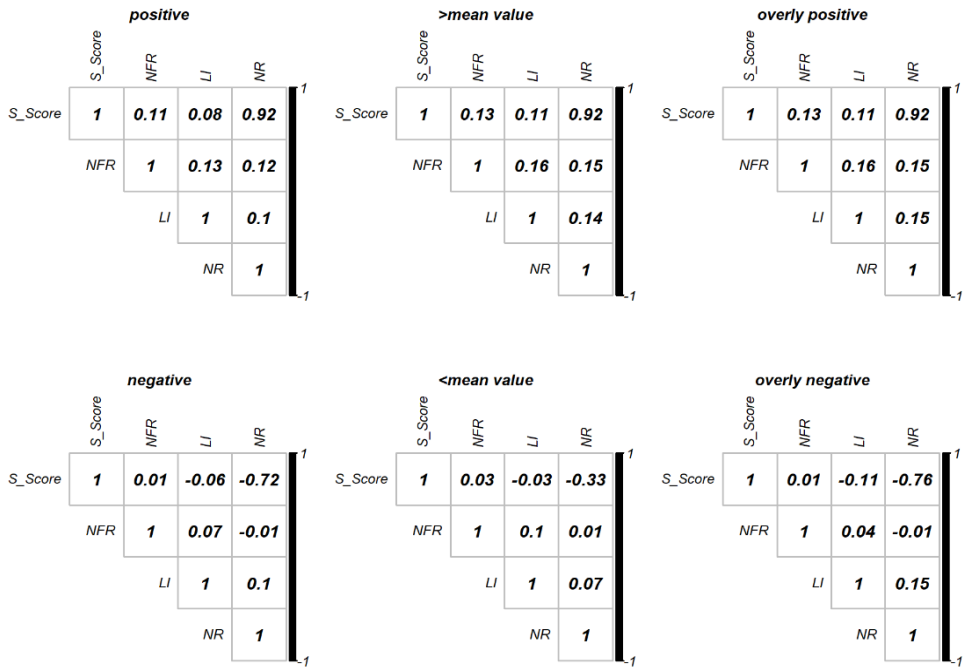

Figure 5. Relationship between user lifespan and biased attitude in the Yelp dataset

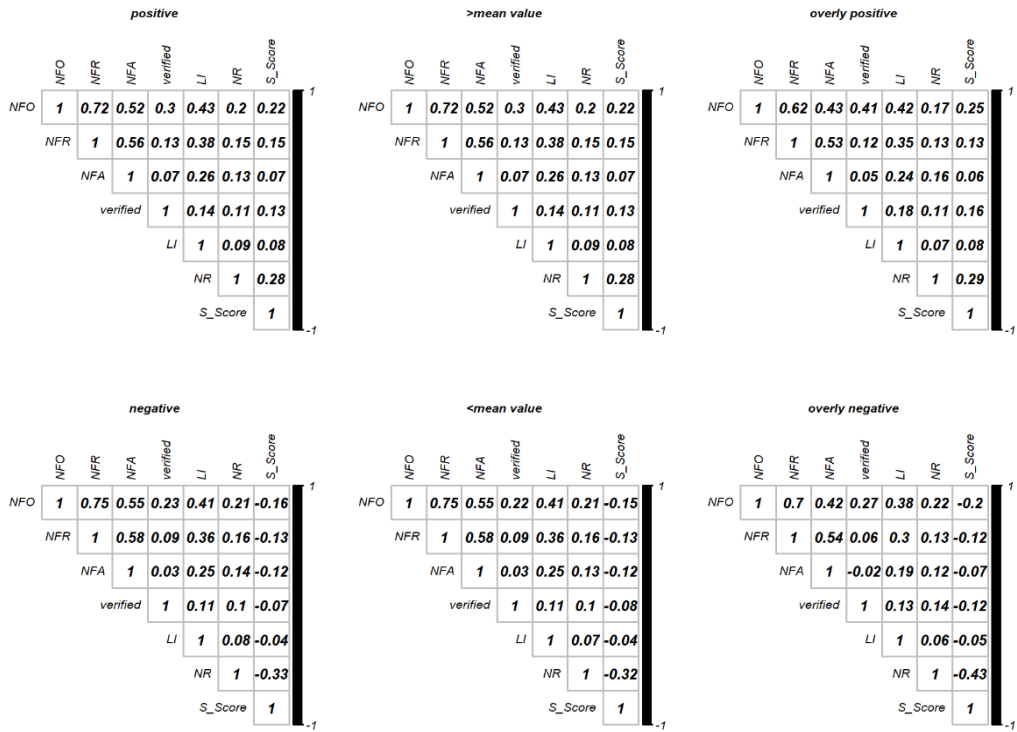

Figure 6. Correlation matrix of COVID-19 tweet texts and user characteristics

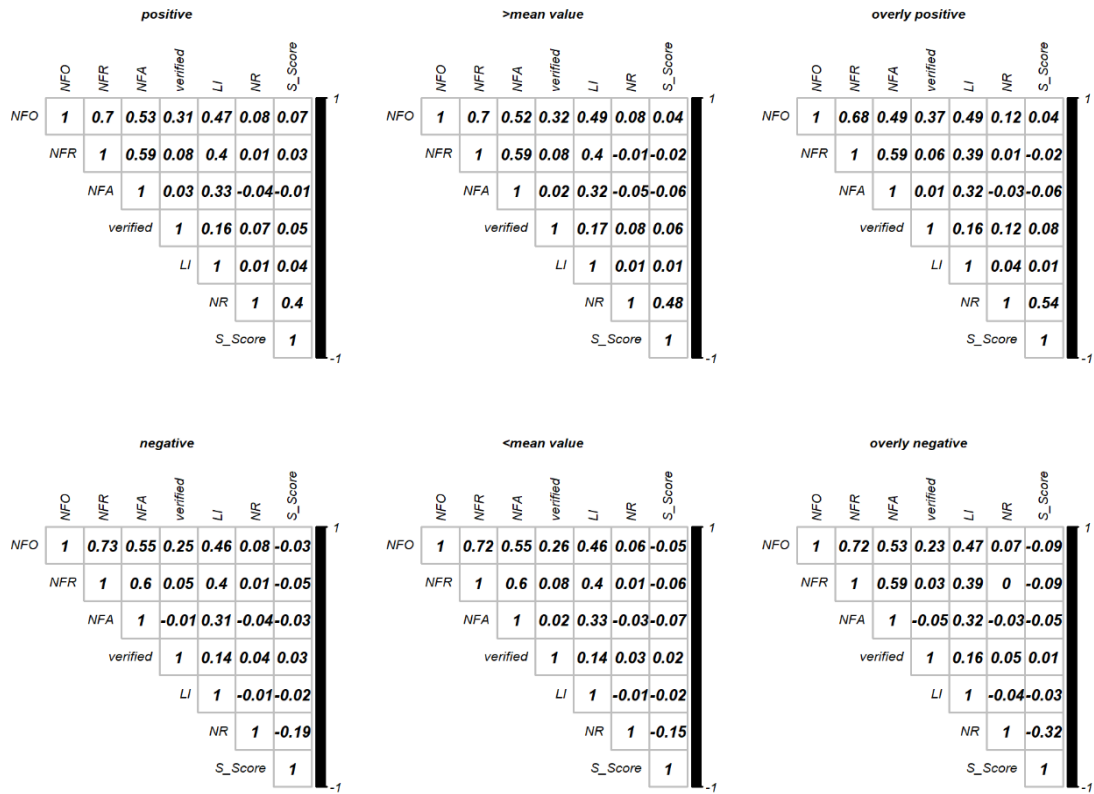

Figure 7. Correlation matrix of COVID-vaccine tweet texts and user characteristics

As illustrated in Figures 5, 6, and 7, there is a relationship between user characteristics and sentiment value. So, it is possible to make a model based on user characteristics to identify biased (overly positive or overly negative) users. In the following, this relationship is described in detail.

*Number of friends and biased attitude:* The number of friends the users have in the network has a significant impact on their own attitude based on the results for the used datasets (i.e., Yelp, COVID-19, and COVID-vaccine). Figure 5 shows that the r value of number of friends and sentiment score is 0.13 for overly positive users in the Yelp dataset. For the COVID-19 dataset (Figure 6), this value for user with sentiment score bigger than the mean value are 0.15 and -0.13 for overly positive and overly negative are 0.13 and -0.12, respectively. For the COVID-vaccine dataset (figure 7), the r value is around -0.1 for overly negative users. These results indicate that the more friends the users have, the higher the tendency that they form a positive attitude and the lower the tendency that they form a negative attitude.

*User lifespan and biased attitude:* Based on prior research (Padel and Foster, 2005; Brucks, 1985), a user's direct experience such as the amount of time the user has spent on the network and the amount of user reviews they have made affect their attitude toward businesses. The duration of membership in social networks is termed the lifespan. The r values of lifespan and sentiment score are 0.11 and -0.11 for overly positive and overly negative users in the Yelp dataset. On the other hand, the r values are 0.08 and -0.05 for the COVID-19 dataset and 0.01 and -0.03 for the COVID-vaccine dataset. One explanation for these results is that when people stay longer in a particular website it is usually because they like that website and they have a positive attitude toward that website. Hence, it makes sense that when we look at those who spent a longer time in the Yelp website that they showed a more positive attitude than a negative attitude.

*Number of reviews and biased attitude:* In order to examine the role of the number of user reviews (direct experience) in user bias, we computed the user sentiment score based on overly positive and overly negative users. Interestingly, we found a positive relationship between the number of user reviews and users with an overly positive attitude, which is known as positivity bias. We also found a negative relationship between the number of user reviews and users with a very negative attitude, which is known as negativity bias. The r values for overly positive and overly negative users are 0.92 and -0.76, 0.29 and -0.46, and 0.54 and -0.32 for the Yelp, COVID-19, and COVID-vaccine datasets, respectively.

*Number of followers and biased attetude.* Another feature of socialization characteristics in social networks is the number of followers. The *r* values for the number of followers and sentiment score for overly positive and overly negative users in the COVID-19 dataset are 0.25 and -0.2, and 0.04 and -0.09 in the COVID-vaccine dataset, respectively. This feature is not available for the Yelp datset.

5.2 Proposed Model

As shown in the previous section, there is a relationship between the user characteristics in the datasets and sentiment value. Therefore, in this section, we present a criterion for the proposed model based on the above-mentioned features. Section 5.1 shows that the number of users' friends (NFR), number of followers (NFO), lifespan (LS) and number of reviews (RN) are important to form the user attitude or UA. First, we normalize these values (NFR, NFO, LS, and RN) by Equation (6):

$$Normal\_value = \frac{x_i - \min(x)}{\max(x) - \min(x)} \qquad (6)$$

After that, a criterion based on the normalized values and the role of each factor in the UA is defined. Significant factors (features) can be derived from the value of the correlation coefficient, principal component analysis, and factor analysis. As shown in the hypothesis development section (5.1), we used the correlation matrix to extract the significant features. In this section based on extracted features, we propose a multi-layer perceptron (MLP) model to identify biased users. Our model labels each user as biased or unbiased based on significant features and the normalized value of each of these features.

One method of machine learning is the MLP neural network. The main parameters of this network are input data, weights, summation and adding bias, activation function and output. The MLP approximates the functional relationships between covariates (inputs) and response variables (outputs).

There are certain input data and each data has some features. For instance, in this study, the features of our data are network structure which depict the psychological behavior of users, such as number of friends, number of followers, lifespan etc. In each step, a data is fed to the network, and based on the weight and bias the network generates an output to a function which is called an activation function, and the output of this function is 0 or 1.

Hence suppose that the MLP network has m features as $X_i, X_2, \ldots, X_m$

$$z = \sum_{i=1}^{m} w_i x_i + bias \qquad (7)$$

This study uses the sigmoid function as follows:

$$\sigma(z) = \frac{1}{1+e^{-z}} \qquad (8)$$

To make a model based on the MLP, we also need to tune the other parameters of this type of neural network as:

$$nn <- neuralnet\ (formula, data, hidden, threshold, rep, startweights, algorithm, err.fct, act.fct, linear.output\ ) \qquad (9)$$

In (9):

*Formula* is the relationship between the covariates and response variables. In our model, we consider it as the relationship between *Biased* and *Number_of_friends, Lifespan, Number_of_Reviews, Number_of_followers*. However, the covariates depend on the user characteristics that are specified in each dataset.

*Data* refers to the dataset used.

*Hidden* refers to a vector that shows the number of hidden layers and hidden neurons in each layer. This study uses trial and error to specify this value. In other words, first we test the different layers and based on the accuracy and error values, we select the best one. From our examination, the number of hidden layer is two, but the number of neurons is different for each dataset.

*Algorithm* refers to the algorithm used to make the neural network. A diverse range of algorithms can be used for this purpose, including *backprop*, *rprop+, rprop-, sag, or slr*. In this study, we use *rprop+*, which means resilient backpropagation and is an improvement on the traditional backpropagation algorithm because the activation function is the sigmoid function; the activation function is affected by the slope of error (Saputra et al., 2017), but *rprop+* uses the sign of the error gradient to update the weight.

*Err.fct* refers to the differentiable error function. Two values can be considered for this parameter, namely, *cross-entropy* and *sum of squared errors (SSE)*. In this study, we use the second one as follows:

$$SSE = \frac{1}{2}\sum_{l=1}^{L}\sum_{h=1}^{H}(O_{lh} - Z_{lh})^2 \quad (10)$$

where $O$ is the predicted output, $Z$ is the observed output, $l = 1 \ldots L$ is the indices of the observation, and $h = 1 \ldots H$ is the output nodes (Günther & Fritsch, 2010). For our problem, $h = 1$ because we have one output node: *Biased*. Therefore

$$SSE = \frac{1}{2}\sum_{l=1}^{L}(O_l - Z_l)^2 \quad (11)$$

*Act.fct* refers to the differentiable activation function. This parameter can be *tanh* and *logistic*. The sigmoid function used in this study is a *logistic* parameter.

$linear.output$ is the final important parameter in Equation (9). If a model does not use activation function, then this parameter is *TRUE,* otherwise it is *FALSE*.

6. EXPERIMENT RESULTS AND DISCUSSION

To measure the performance of a model, two methods are usually used: 1) evaluation based on the assumptions to which the model should apply and 2) evaluation based on the efficiency of the model in predicting new values (not observed). We evaluated our model using the former because this study assumed that physiological features can form the UA and that bias in users is related to psychological behavior based on existing psychological studies mentioned in section 5

In order to demonstrate the accuracy of the proposed model, we used the neural network to determine the accuracy of the proposed model based on the above-mentioned user characteristics (i.e., number of friends, lifespan, number of reviews, and number of followers). Tables 4, 5, and 6 present the results of the proposed neural network model based on the criterion defined in Equation (9) for the Yelp, COVID-19, and COVID-vaccine datasets, respectively, together with the accuracy calculations.

Table 4. Contingency matrix of proposed model for Yelp dataset

| Observed | Predicted | |
|---|---|---|
| | Neg | Pos |
| Neg | 99.8 | 0.2 |
| Pos | 21.3 | 78.7 |

$$Accuracy = \frac{TP + TN}{TP + TN + FP + FN} = \frac{78.7 + 99.8}{78.7 + 99.8 + 0.2 + 21.3} = 89\%$$

Table 5. Contingency matrix of proposed model for COVID19

| Observed | Predicted | |
|---|---|---|
| | Neg | Pos |
| Neg | 66.5 | 33.5 |
| Pos | 33.2 | 66.8 |

$$Accuracy = \frac{TP + TN}{TP + TN + FP + FN} = \frac{66.5 + 66.8}{66.5 + 66.8 + 33.5 + 33.2} = 67\%$$

Table 6. Contingency matrix of proposed model for Covid vaccine

| Observed | Predicted | |
|---|---|---|
| | Neg | Pos |
| Neg | 99.8 | 0.2 |
| Pos | 37.8 | 62.2 |

$$Accuracy = \frac{TP + TN}{TP + TN + FP + FN} = \frac{62.2 + 99.8}{62.2 + 99.8 + 0.2 + 37.8} = 81\%$$

The experiment results show that the accuracy rates of the proposed model when applied to the Yelp, COVID-19, and COVID-vaccine datasets are 89%, 67%, and 81%, respectively. The results indicate that biased users can be predicted accurately based on their psychological behavior. Figure 8 shows the proposed neural network, from which it can be seen that the covariates of the COVID-19 and COVID-vaccine datasets are number of followers, number of friends, lifespan, and number of reviews, while the input data for the neural network of the Yelp dataset are number of friends, lifespan, and number of reviews.

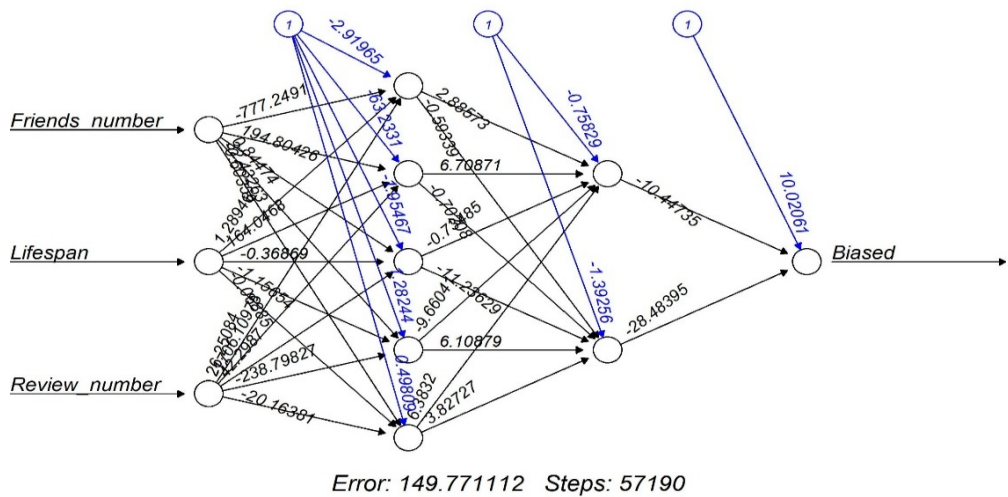

(a)

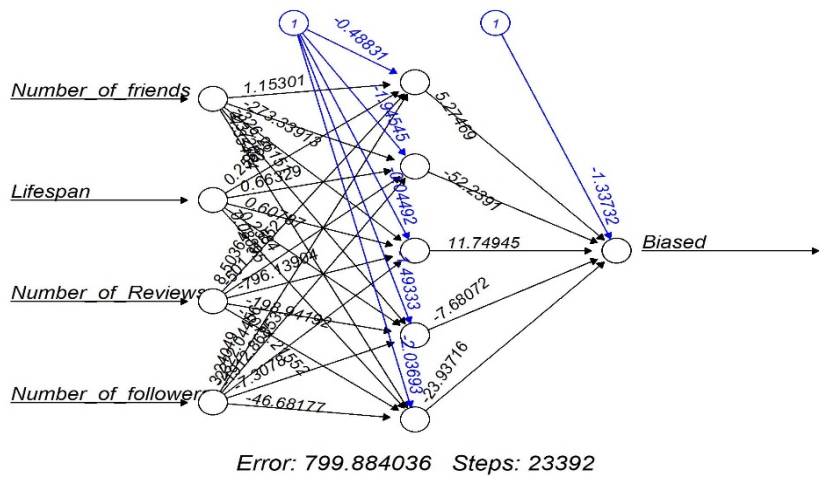

(b)

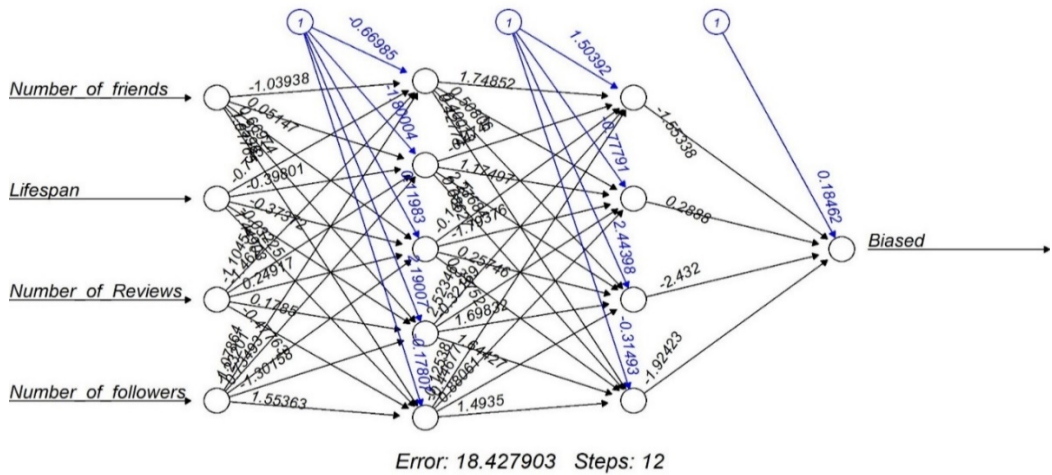

(c)

Figure 8. Neural network based on user behavior to predict biased users' attitudes for (a) Yelp, (b) COVID-19, and (c) COVID-vaccine dataset

Figure 9 shows the weighting of each user behavior characteristic for the neural network model. The result demonstrates that the user characteristics have a significant effect on the proposed model.

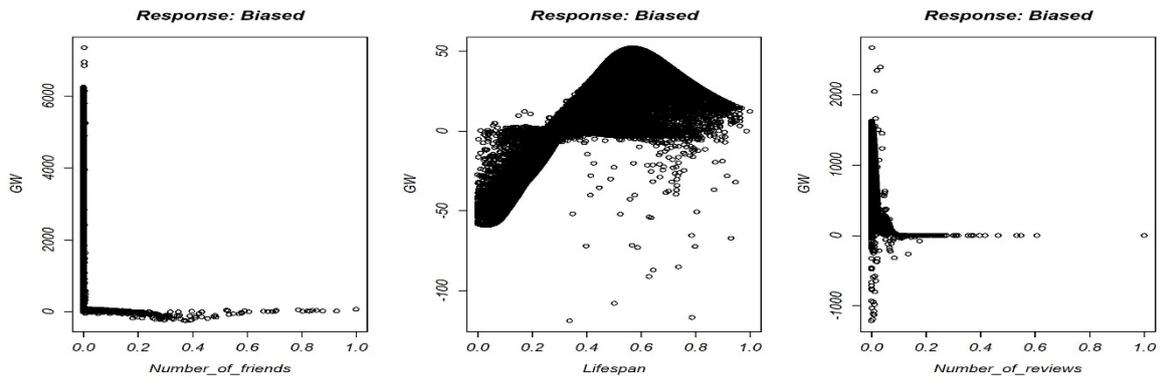

(a)

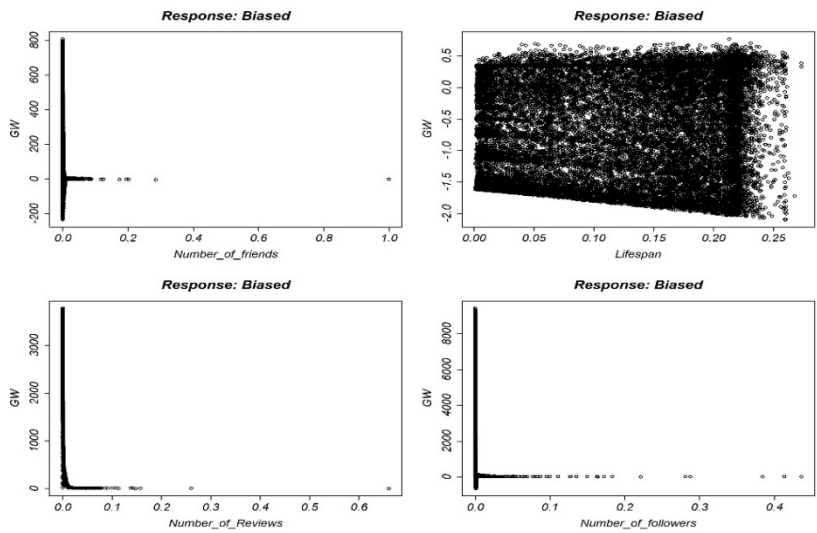

(b)

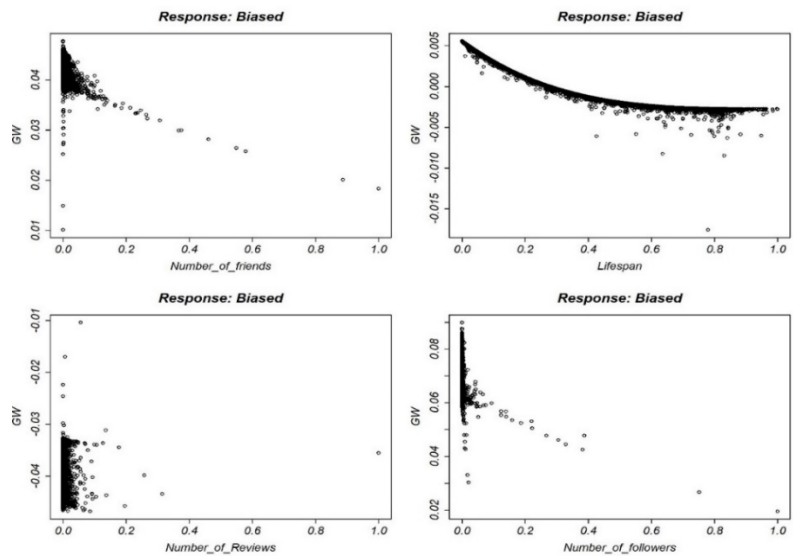

(c)

Figure 9. Weight (GW) of each user behavior characteristic for the proposed neural network model to identify biased users' attitudes in (a) Yelp, (b) COVID-19, and (c) COVID-vaccine dataset

7. CONCLUSION

This study proposed a MLP neural network classifier based on user characteristics to identify biased users by their psychological behavior. The results of an experiment conducted on three different datasets showed that the user characteristics that indicate the users' psychological behavior are influential in user attitude polarities. The number of friends and number of followers show the users' socialization, while the lifespan and number of reviews show the users' direct experience. Socialization affects users' sentiments because people who are socialized are more positive than isolated people in OSNs. In addition, knowledge and subsequent experience acquisition, which is implicit in the number of reviews made by users in the datasets can affect the strength of user attitude polarity. We demonstrated that as a user gains more direct personal experience, the strength of positive polarity increased, while conversely, the strength of negative polarity decreased. This study paves the way for a new research direction in which user review text cannot be used as the sole indicator when analyzing business or product quality because user attitude also plays an important role in such a judgment. It seems clear from this study that user positivity and negativity may affect user judgments of businesses. Hence biased attitudes should be removed for unbiased sentiment analysis. Above all, this study shows that computers can utilize user psychological variables through the social network structure to predict user's attitude without literally analyzing users' review texts or posts.